\documentclass[aps,twocolumn,showpacs,superscriptaddress,amsmath,amssymb,amsfonts,floatfix,longbibliography]{revtex4-2}

\usepackage{array}[=2016-10-06]
\usepackage[T1]{fontenc} 
\usepackage{graphicx}
\usepackage{dcolumn}
\usepackage{bm}
\usepackage{amssymb}
\usepackage{tabularx}
\usepackage{microtype}
\usepackage{xfrac}
\usepackage{chemformula}
\usepackage{gensymb}
\usepackage[most]{tcolorbox}
\usepackage{xcolor}
\usepackage{multirow}
\usepackage{booktabs}
\usepackage[flushleft]{threeparttable}
\usepackage{makecell}

\usepackage{hyperref}

\hypersetup{
	colorlinks=true, %set true if you want colored links
	citecolor=blue,  %choose some color if you want links to stand out
}

\begin{document}

\title{Machine Learning Guided Discovery of Corundum High Entropy Oxides}

\author{Abraham A. Mancilla}
      \affiliation{Stewart Blusson Quantum Matter Institute, University of British Columbia, Vancouver, BC V6T 1Z4, Canada}
    \affiliation{Department of Physics \& Astronomy, University of British Columbia, Vancouver, BC V6T 1Z1, Canada}

\author{Oliver A. Dicks}
    \affiliation{Stewart Blusson Quantum Matter Institute, University of British Columbia, Vancouver, BC V6T 1Z4, Canada}
    \affiliation{School of Physical and Chemical Sciences, Queen Mary University of London, Mile End Road, London E1 4NS, United Kingdom}

\author{Solveig S. Aamlid}
    \affiliation{Stewart Blusson Quantum Matter Institute, University of British Columbia, Vancouver, BC V6T 1Z4, Canada}
    \affiliation{Laboratoire de Physique de la Matière Condensée, CNRS, École polytechnique, Institut Polytechnique de Paris, 91120 Palaiseau, France}

\author{Mario Ulises Gonz\'alez-Rivas}
    \affiliation{Stewart Blusson Quantum Matter Institute, University of British Columbia, Vancouver, BC V6T 1Z4, Canada}
    \affiliation{Department of Physics \& Astronomy, University of British Columbia, Vancouver, BC V6T 1Z1, Canada}
    
\author{Karl Tsang}
    \affiliation{Stewart Blusson Quantum Matter Institute, University of British Columbia, Vancouver, BC V6T 1Z4, Canada}
\affiliation{Department of Chemistry, University of British Columbia, Vancouver, BC V6T 1Z1, Canada}

\author{Dongjoon Song}
    \affiliation{Stewart Blusson Quantum Matter Institute, University of British Columbia, Vancouver, BC V6T 1Z4, Canada}
    
\author{J\"org Rottler}
    \affiliation{Stewart Blusson Quantum Matter Institute, University of British Columbia, Vancouver, BC V6T 1Z4, Canada}
    \affiliation{Department of Physics \& Astronomy, University of British Columbia, Vancouver, BC V6T 1Z1, Canada}

\author{Alannah M. Hallas}
\email[Email: ]{alannah.hallas@ubc.ca}
    \affiliation{Stewart Blusson Quantum Matter Institute, University of British Columbia, Vancouver, BC V6T 1Z4, Canada}
    \affiliation{Department of Physics \& Astronomy, University of British Columbia, Vancouver, BC V6T 1Z1, Canada}
    \affiliation{Department of Chemistry, University of British Columbia, Vancouver, BC V6T 1Z1, Canada}
    \affiliation{Canadian Institute for Advanced Research (CIFAR), Toronto, ON, M5G 1M1, Canada}

\date{\today}% It is always \today, today,

\begin{abstract}

Early thinking in the field of high entropy oxides (HEOs) emphasized their likely abundance, with combinatorial arguments hinting at a myriad of new materials. The experimental reality has proven more challenging: the stability of HEOs cannot be straightforwardly predicted based on ionic radii, lattice geometry, and charge-balancing considerations alone. In this work, we employ machine learning interatomic potentials (MLIPs) to predict the synthesizability of HEOs of the form $A_2$O$_3$ derived from a selection of trivalent cations. From nearly 500 possible compositions, we identify 16 promising candidates for experimental validation with solid-state and combustion synthesis. We discover three new HEOs in the corundum structure, including (Al,Cr,Fe,Rh,Sc)$_2$O$_3$, and one novel cation-ordered phase, (Al,Fe,Ga,Sc)$_2$O$_3$. By far the most common synthesis outcome was a mixture of competing phases, sometimes involving redox reactions. Our results also reveal profound synthesis method dependence for the final product, where qualitatively equivalent outcomes between the two synthesis methods were only observed for 3 of the 16 tested compositions. We conclude that the occurrence rate of HEOs is far rarer than initially believed and that machine learning approaches can effectively guide us to the ``needle in the haystack''.

\end{abstract}

\maketitle

\section{Introduction}

Following their initial discovery in 2015~\cite{Rost2015}, early thinking in the field of high entropy oxides (HEOs) emphasized their likely abundance. Indeed, the vast combinatorial space in which to design materials where large numbers of cations share a single sublattice appears, at first, nearly infinite. Part of this intuition originated from the alloys field, where mixtures of even 10-12 metallic elements can form a single phase~\cite{cantor_thermodynamics_2025}. However, the experimental reality in oxides is more complex. Whereas metallic bonding in alloys combined with their relatively simple crystal structures provides an ideal platform for solid solutions, the covalent and ionic bonding of oxides, coupled with their more elaborate crystal structures, yields a more limited range of stability. %Furthermore, cations sharing a sublattice must have compatible ionic radii and oxidation states, further restricting the pool of possible materials. 

Taking rock salt-type HEOs as an example, we can quickly see how the earlier combinatorial promise dwindles. The 65 choose 5 (8 million) ways to select from radioactively stable metals are first reduced to 27 choose 5 (80 thousand) to represent the number of elements that can accommodate a stable divalent oxidation state. Finally, there are further limitations according to oxygen chemical potential and ionic radii compatibility. The end result is ultimately a small handful of synthesizable materials requiring tight control of synthesis conditions. Rock salt-type HEOs have been intensively studied, starting from the prototype material \ch{(Mg,Co,Ni,Cu,Zn)O}~\cite{Rost2015}, which has since been expanded to a variety of Mn and Fe substituted analogs by careful control of oxidation conditions during synthesis~\cite{Lin2020,pu_mgmnfeconio_2023,almishal_thermodynamics-inspired_2025}. 
%to the Mn and Fe substituted compounds prepared by mechanochemical synthesis under argon \cite{Lin2020}, \ch{(Mg,Mn,Fe,Co,Ni)O} produced in inert atmosphere \cite{pu_mgmnfeconio_2023}, and most recently a new handful of mixed Mn/Fe compounds made under oxygen-poor conditions \cite{almishal_thermodynamics-inspired_2025}.
 
\begin{figure*}[htbp]
\centering
\includegraphics[width=\textwidth]{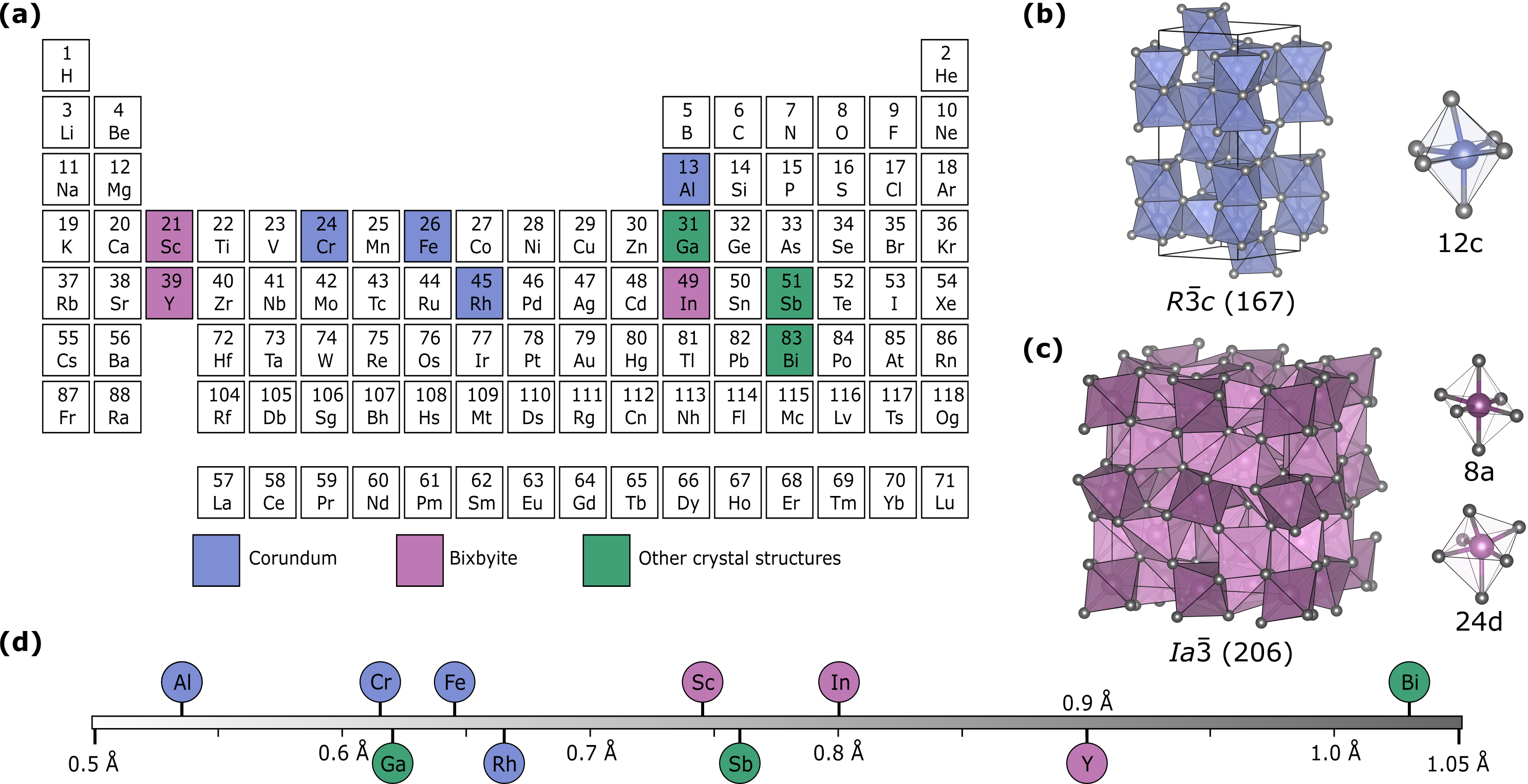}
 \caption{\textbf{Design considerations for a trivalent high entropy oxide.} \textbf{(a)} Periodic table showing the considered metals with 3+ as their most stable oxidation state, shaded with the corresponding color for the structure: blue for corundum, pink for bixbyite, and green for \ch{Ga_2O_3}, \ch{Bi_2O_3}, and \ch{Sb_2O_3}, which crystallize in three different crystal structures. \textbf{(b)} Trigonal unit cell of the corundum structure. %found for \ch{Cr_2O_3}, \ch{Al_2O_3}, \ch{Fe_2O_3}, and \ch{Rh_2O_3}. 
 The single slightly distorted octahedral environment for the metal cations in Wyckoff site $12c$ is shown on the right. \textbf{(c)} The cubic unit cell of the bixbyite structure. % found for \ch{Sc_2O_3}, \ch{Y_2O_3}, and \ch{In_2O_3}. 
 The two pseudo-octahedral environments for the metal cations in Wyckoff sites $8a$ and $24d$ are shown on the right. \textbf{(d)} The selected cations are organized according to their ionic radii and shaded with the color corresponding to their structure.}
\label{fig:periodictable}
\end{figure*}

The combinatorial complexity of the HEO search space, combined with their rare occurrence rate, calls for a more targeted approach to accelerate the discovery of new compounds. Computational screening therefore becomes desirable and has been successfully applied to high entropy %alloys~\cite{lederer_search_2018}, 
ceramics~\cite{sarker2018high,divilov_disordered_2024}, including oxides~\cite{Pitike2022,aamlid2023understanding}. However, customary density functional theory (DFT) methods, whilst accurate, are limited in scope due to the computational cost associated with the large unit cells required to accurately represent the cation disorder in an HEO and the large compositional search space. Early DFT descriptor-based approaches that approximated the full HEO disorder by averaging over 2-cation systems demonstrated predictive capability~\cite{Pitike2022,aamlid2023understanding} with reduced computational cost, but at the expense of reducing accuracy and the number of compounds that could be screened. Recent developments in universal machine learning interatomic potentials (MLIPs) have enabled the calculation of mixing enthalpies and disorder descriptors for large, fully disordered HEO unit cells at a fraction of the cost of ab initio methods while retaining near DFT-level accuracy~\cite{sivak_discovering_2025,dicks_expanding_2026}. %This was demonstrated for the rocksalt HEOs~\cite{sivak_discovering_2025} using CHGNet~\cite{deng_chgnet_2023}, correctly identifying known synthesized HEOs out of approximately 100 compositions. Recently, we extended this approach to thousands of tetravalent $A\text{O}_2$ systems with 14 elements and 3 crystal structures in 4- and 5-component HEOs, introducing a new entropy descriptor derived from the individual atomic energies only accessible from MLIPs~\cite{dicks_expanding_2026} which correctly distinguished the single known phase. Ideally, the temperature and oxygen partial pressure dependent free energies of HEOs could be calculated against all competing phases on the hull, but full thermodynamic integration remains intractable if the goal is to screen thousands of compounds. 

In this article, we describe a combined computational and experimental approach to the discovery of new HEOs. To test this strategy, we identify a significant gap among known HEOs corresponding to the trivalent binary oxide form, $A_2$O$_3$. We use chemical arguments to define a search space of 10 stable trivalent cations, yielding nearly 500 possible four- and five-component HEOs. Leveraging MLIPs, we efficiently survey this large compositional space and use entropy- and enthalpy-based descriptors to identify the 16 compositions most likely to form stable phases. Experimental validation leads to the discovery of 3 new HEOs in the corundum structure, while the majority of synthesis attempts result in multiphase mixtures. Taken together, these results suggest that the true occurrence rate of HEOs is significantly lower than initially believed, and demonstrate that machine learning approaches can effectively guide discovery within an otherwise intractable compositional landscape.

\section{Design of trivalent HEOs}

Stable 3+ valence states are arguably the most prevalent oxidation state represented on the periodic table. In the realm of high entropy, however, the only reported trivalent oxides of the form $A_2$O$_3$ %with a single cation lattice 
are those in which the cation site is occupied by multiple rare earths~\cite{djenadic2017multicomponent,heo-bixbyite1}. These mixtures, which inhabit a bixbyite structure, are favored for solid solution formation due to the ``pronounced chemical similarities'' of the rare earth elements~\cite{cheisson2019rare}; thus, while they can have large configurational entropies, they are likely not entropy-stabilized~\cite{aamlid2023understanding}. Meanwhile, the numerous trivalent transition metals and main group elements are not yet meaningfully incorporated in any $A_2$O$_3$-type trivalent HEO. %The inclusion of these elements is also essential if one wishes to achieve entropy stabilization. I think calling it a binary trivalent HEO does not make sense, as binary should mean composed of oxygen and one cation -SSAa.
This phase space is therefore ripe for computational and experimental exploration.

%With the exception of trivial mixtures of rare earths, no binary trivalent HEOs of the form $A_2$O$_3$ are yet known. 

%As cations with stable trivalent oxidation states are among the most common in the periodic table, the absence of a trivalent $A_2$O$_3$ HEO is notable and unexpected.

To enable our computational screening, our first step is to establish which elements and structures we are going to consider. We limit our scope to include cations for which 3+ is the overwhelmingly favored oxidation state and we exclude the lanthanide elements to avoid trivial bixbyite phases. These criteria lead us to the selection of 10 cations spanning the periodic table, which are graphically represented in Fig.~\ref{fig:periodictable}(a). These include the two light rare earths (Y and Sc), three transition metals (Cr, Fe, and Rh), three boron group elements (Al, Ga, and In), and two pnictogens (Sb and Bi).
As experimental validation is an intended aspect of this study, we chose to exclude thallium and arsenic due to their toxicity. Other cations (such as Mn, Co, and Ru, among others) were also excluded due to their propensity to adopt other oxidation states, complicating both their stability prediction with MLIPs and their experimental validation.

The highlighted elements in Fig. \ref{fig:periodictable}(a) are color-coded according to their most stable binary oxide crystal structure type under ambient conditions. The corundum structure (Fig. \ref{fig:periodictable}b) % named after the corundum mineral $\alpha$-\ch{Al_2O_3}, 
is the most prevalent among our selected elements. %It is a stable polymorph at ambient conditions for half of the elements considered in this study. 
This trigonal structure, with space group \textit{R$\overline{3}$c} (No. 167), consists of alternating close-packed layers of oxygen ions and metallic cations, in which the cations fill two-thirds of the octahedral voids. The cation site has a distorted octahedral oxygen coordination, and these octahedra form face-, edge-, and corner-sharing motifs. %Being one of the most prevalent structures in the trivalent system and a single cation phase, made the corundum structure a strong candidate to host a trivalent HEO.
The second most common structure for our trivalent cations is the bixbyite structure. %, which is also shared by the heavier lanthanides. %, is named after the bixbyite mineral \ch{(Mn,Fe)_2O_3}. 
%It is a stable polymorph at ambient conditions for around a third of the elements considered in this study. 
This cubic structure has space group \textit{Ia}$\overline{3}$ (No. 206) and contains two crystallographically unique pseudo-octrahedral cation sites, which are shaded darker ($8c$) and lighter ($24d$) in Fig.~\ref{fig:periodictable}(c). %, are both pseudo-octrahedral with $24d$ being more distorted. %It is a two-cation site crystal structure with both crystallographic sites in an pseudo-octahedrally coordinated environment of oxygen anions, but the two sites do have slight variations. The two sites are labeled with the Wyckoff positions 24\textit{d} and 8\textit{b}, the 8\textit{b} site is an octahedra with equal bond lengths but distorted angles between bonds, and the 24\textit{d} site has three distinct bonnd lengths, and a more pronounced distortion in the angle between bonds. 
%Notably, it is also a metastable structure of \ch{Fe_2O_3}, but it is scarce and transforms easily into corundum at high temperatures\cite{bixbyite_iron}.  %Being one of the most prevalent structures in the trivalent system and having two cation sites that are similar enough, made the bixbyite structure a strong alternative to corundum to host trivalent HEO.
The remaining three elements each occupy a unique binary oxide crystal structure. Under ambient conditions, the most stable polymorph of Ga$_2$O$_3$ is the monoclinic $\beta$-Ga$_2$O$_3$ phase, where Ga is found in both tetrahedral and octahedral coordination. Lastly, Sb$_2$O$_3$ and Bi$_2$O$_3$, both exhibit crystal structures with highly anisotropic local oxygen coordinations, which can be attributed to their lone pairs.
%We therefore focus our search on corundum and bixbyite structures, as they are overwhelmingly the most common for trivalent cations while the alternative crystal structures are found uniquely for a single binary oxide. %We therefore direct our search within these two crystal structures. 

Besides structure and local coordination, a third important factor for solid solution formation is ionic radius as outlined in the Hume-Rothery rules. In Fig.~\ref{fig:periodictable}(d), the 10 selected trivalent cations are ranked according to their ionic radius and color-coded according to their structure. The specified radius corresponds to the trivalent six-fold coordinate effective ionic radius as defined by Shannon~\cite{shannon1976revised}. This highlights that the corundum structure is favored for the smaller cations, while bixbyite is favored for larger ionic radii. Our ensemble of cations span a large range of radii, from the smallest cation Al at 0.54 \AA\ to the largest cation Bi at 1.03 \AA. Seven of the ten cations have ionic radii between 0.6 and 0.8~\AA.

The 10 cation phase space considered here yields 210 possible four-component and 252 possible five-component systems. Certain favorable cation combinations are immediately evident by consideration of their shared crystal structure and similar ionic radii, such as (Al,Cr,Fe,Rh)$_2$O$_3$,  %(Al,Cr,Fe,Ga,Rh)$_2$O$_3$, 
all of which independently inhabit a corundum structure. However, %as we will see, 
such a simplistic approach does not uniquely identify synthesizable materials, and it excludes other materials that could form. Moreover, previous experimental work has demonstrated that ionic radius alone is not predictive of phase stability in HEOs~\cite{khoroshun2025more}. For the 462 possible four- and five-component materials considered here, a brute force experimental approach is impractical (and becomes even more impractical due to the combinatorial nature of these compositions or with the introduction of non-equimolar compositions). We therefore turn to MLIPs to help narrow down this large compositional space.  

%Antimony and bismuth were the two element that did not have either corundum or bixbyite as a stable polymorph. Antimony oxide has two different stable  polymorphs, the most stable at lower temperatures  

\begin{table*}
    \caption{The 16 compositions selected for experimental validation and summary of outcomes with solid state and combustion synthesis.}
\centering
\begin{tabular}{lc>{\centering\arraybackslash}p{1.9cm}>{\centering\arraybackslash}p{1.6cm}>{\centering\arraybackslash}p{1.4cm}>{\centering\arraybackslash}p{1.9cm}>{\centering\arraybackslash}p{2.9cm}>{\centering\arraybackslash}p{2.9cm}}
\hline
\multicolumn{1}{l}{\multirow{2}{*}{\begin{tabular}[c]{@{}c@{}}Composition \\ \end{tabular}}} 
& \multicolumn{1}{c}{\multirow{2}{*}{\begin{tabular}[c]{@{}c@{}}Label \\ \end{tabular}}} 
& \multicolumn{1}{c}{\multirow{2}{*}{\begin{tabular}[c]{@{}c@{}}$\Delta H_{\text{MACE}}$ \\ $\left(\sfrac{\text{eV}}{\text{f.u.}}\right)$\end{tabular}}} 
& \multicolumn{1}{c}{\multirow{2}{*}{\begin{tabular}[c]{@{}c@{}}$\sigma_{E_i}$ \\ (eV)\end{tabular}}}   
& \multicolumn{1}{c}{\multirow{2}{*}{\begin{tabular}[c]{@{}c@{}}$\rho$ \\ \end{tabular}}} 
& \multicolumn{1}{c}{\multirow{2}{*}{\begin{tabular}[c]{@{}c@{}}Predicted \\ Structure\end{tabular}}} 
& \multicolumn{1}{c}{\multirow{2}{*}{\begin{tabular}[c]{@{}c@{}}Result of \\ Solid State \end{tabular}}} 
& \multicolumn{1}{c}{\multirow{2}{*}{\begin{tabular}[c]{@{}c@{}}Result of \\ Combustion\end{tabular}}}
\\
\\ \hline
\multicolumn{8}{c}{Four-component} \\ \hline
\ch{(Al,Cr,Fe,Ga)2O3}$^{*}$ & 4c-1 & 0.046 & 0.041 & 0.276 & corundum & Mixture of phases & Phase pure HEO \\ \hline
\ch{(Al,Cr,Fe,Sc)2O3}& 4c-2 & 0.147 & 0.054 & 0.429 & corundum  & Mixture of phases & Mixture of phases \\ \hline
\ch{(Al,Fe,Ga,Sc)2O3}& 4c-3 & 0.134 & 0.060 & 0.453 & corundum & Ordered phase & Ordered phase \\ \hline
\ch{(Cr,Fe,Ga,In)2O3}& 4c-4 & 0.138 & 0.053 & 0.418 & bixbyite & Mixture of phases & Mixture of phases \\ \hline
\ch{(Cr,Fe,Ga,Rh)2O3} & 4c-5 & 0.080 & 0.054 & 0.378 & corundum & Redox reaction & Not attempted \\ \hline
\ch{(Cr,Fe,Ga,Sc)2O3} & 4c-6 & 0.088 & 0.059 & 0.414 & bixbyite & Mixture of phases & Mixture of phases \\ \hline
\ch{(Cr,Fe,In,Sc)2O3} & 4c-7 & 0.142 & 0.060 & 0.462 & bixbyite & Mixture of phases & Mixture of phases \\ \hline
\ch{(Cr,Fe,Rh,Sc)2O3} & 4c-8 & 0.110 & 0.067 & 0.474 & corundum & Redox reaction & Not attempted \\ \hline
\multicolumn{8}{c}{Five-component} \\ \hline
\ch{(Al,Cr,Fe,Ga,In)2O3} & 5c-1 & 0.199 & 0.054 & 0.486 & bixbyite & Mixture of phases & Mixture of phases \\ \hline
\ch{(Al,Cr,Fe,Ga,Rh)2O3}$^{**}$ & 5c-2 & 0.119 & 0.055 & 0.410 & corundum & Redox reaction & Redox reaction \\ \hline
\ch{(Al,Cr,Fe,Ga,Sc)2O3} & 5c-3 & 0.134 & 0.051 & 0.403 & corundum & Mixture of phases & Mixture of phases \\ \hline
\ch{(Al,Cr,Fe,Rh,Sc)2O3} & 5c-4 & 0.200 & 0.062 & 0.543 & corundum & Phase pure HEO & Not attempted \\ \hline
\ch{(Cr,Fe,Ga,In,Rh)2O3} & 5c-5 & 0.186 & 0.065 & 0.533 & bixbyite & Mixture of phases & Not attempted \\ \hline
\ch{(Cr,Fe,Ga,In,Sc)2O3} & 5c-6 & 0.160 & 0.063 & 0.493 & bixbyite & Mixture of phases & Mixture of phases \\ \hline
\ch{(Cr,Fe,Ga,Rh,Sc)2O3} & 5c-7 & 0.127 & 0.056 & 0.422 &  corundum & Phase pure HEO & Not attempted \\ \hline
\ch{(Cr,Fe,Ga,Sc,Y)2O3}$^{***}$ & 5c-8 & 0.223 & 0.065 & 0.564 & bixbyite & Mixture of phases & Mixture of phases \\ \hline
\end{tabular}
    \\{\raggedright \footnotesize $^{*}$: compound with the lowest stability descriptor $\rho$ out of all 462 compositions. \\
    $^{**}$: five-component compound with the lowest stability descriptor $\rho$ out of all 252 five-component compositions.\\
    $^{***}$: Lowest enthalpy composition containing Y, which fell outside the defined cutoff.\\}
    \label{tab:Summaryattempts}
\end{table*}

\section{Computational structure search}

\begin{figure}[htbp]
  \begin{center}
\includegraphics[width=\columnwidth]{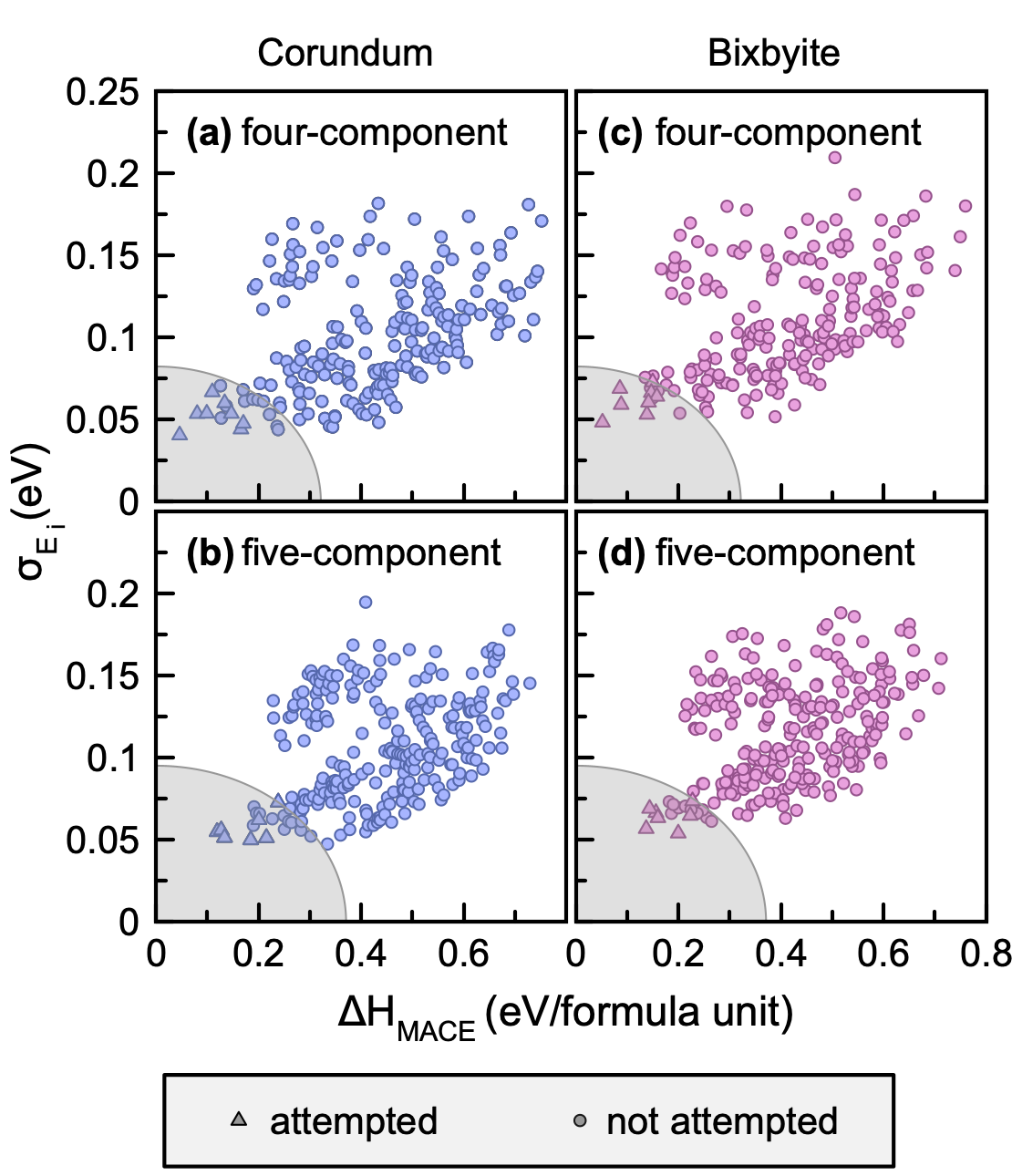}      \caption{\textbf{Stability parameters of trivalent high entropy oxides calculated using MLIPs.} The markers represent the compositions, where their location is represented by the enthalpy of mixing $\Delta H_{\text{MACE}}$ in the $x$-axis, and the entropy descriptor $\sigma_{E_i}$ in the $y$-axis. The triangle markers are the compositions selected for experimental validation, and the circles are compositions that were not attempted. The left panels \textbf{(a,b)} with the blue markers are for the calculations done in the corundum structure, and the right plots \textbf{(c,d)} with the pink markers are for the calculations done in the bixbyite structure. The plots on the top \textbf{(a,c)} represent the calculations done for four-component compounds, and the bottom \textbf{(b,d)} for the five-component compounds. The gray shaded area represents the area of highest anticipated synthesizability, and defines the cutoff for our selection.}
      \label{fig:MLIPs}
  \end{center}
\end{figure}

In order to pare down the 462 possible four- and five-component high entropy oxide compounds to a targeted, practicable list for experimental synthesis attempts we used a computational structure search method previously tested on tetravalent compounds~\cite{dicks_expanding_2026}. Each compound was screened using MLIPs in both corundum and bixbyite crystal structures, yielding 924 relaxed structures in total. From each we extracted 2 descriptors, the enthalpy of mixing $\Delta H_{\text{MACE}}$ and an entropy descriptor $\sigma_{E_{i}}$.

The enthalpy of mixing, $\Delta H_{\text{MACE}}$, is defined here as the enthalpy cost of forming the HEO with respect to its constituent binary oxides on the enthalpy hull. The lower $\Delta H_{\text{MACE}}$, the lower the temperature at which the configurational entropy contribution to the free energy ($-k_{\text{B}}T\text{ln}(n_{\text{cation}})$) stabilizes the disordered HEO phase. A lower $\Delta H_{\text{MACE}}$ also means there is a lower likelihood of competing ternary oxide phases, which may have lower free energies at high temperature. The entropy descriptor, $\sigma_{E_{i}}$, quantifies the degree of homogeneity of the local cation environment across the relaxed supercell, similar to the entropy forming ability descriptors previously introduced for carbides~\cite{sarker2018high}. A low $\sigma_{E_{i}}$ indicates all the cations in the HEO have similar, stable surroundings regardless of the configuration of neighbouring elements, indicative of a random solid solution whereas a high  $\sigma_{E_{i}}$ suggests the system is likely to favour phase separation or local ordering. 

The distribution of $\Delta H_{\text{MACE}}$ against $\sigma_{E_{i}}$ is shown in Fig.~\ref{fig:MLIPs} for all compositions screened. The top panels show compositions with four-components while the bottom panels show compositions with five-components in the corundum (left) and bixbyite (right) structures, respectively. The overall distribution of the data points is similar, independent of structure type and constituent number and larger $\Delta H_{\text{MACE}}$ values generally correlate with a larger $\sigma_{E_{i}}$. When examining the distribution of the descriptors for the two structure types, the average $\Delta H_{\text{MACE}}$ values are 432 and 411 meV/formula unit for corundum and bixbyite respectively, where-as corundum has a lower mean $\sigma_{E_{i}}$ of 104 meV compared to 112 meV for bixbyite.  The distribution of $\Delta H_{\text{MACE}}$ values at both the low and high end is notably broader for the four-component compositions than in the five-component case, consistent with a regression to the mean as more constituents are included. This screening reveals that the majority of compositions have high values of both descriptors, consistent with the empirical rarity of HEOs despite the number of possibilities implied by simple combinatorics.

To maximize the likelihood of successful synthesis, we additionally introduce a thermodynamic stability descriptor, $\rho$ (see Eqn.~\ref{eq:MEED}), which is a weighted optimization of $\Delta H_{\text{MACE}}$ and $\sigma_{E_{i}}$~\cite{MLIPS_tetravalent}. The gray region in Fig. \ref{fig:MLIPs} colors all points that lie below the highest $\rho$-value of the attempted compositions (see Table \ref{tab:Summaryattempts}). This descriptor additionally allows us to identify the most probable crystal structure type for each HEO composition. Across the 462 compositions considered here, we find that bixbyite is the preferred structure in 341 cases, amounting to 74\% of all compositions, in contrast with just 121 compositions that favor corundum. The skewed preference towards bixbyite is primarily driven by the larger ionic radii cations, and the greater structural flexibility of the bixbyite local oxygen environment. 

\section{Experimental validation}

\begin{figure*}[htbp]
\centering
\includegraphics[width=\textwidth]{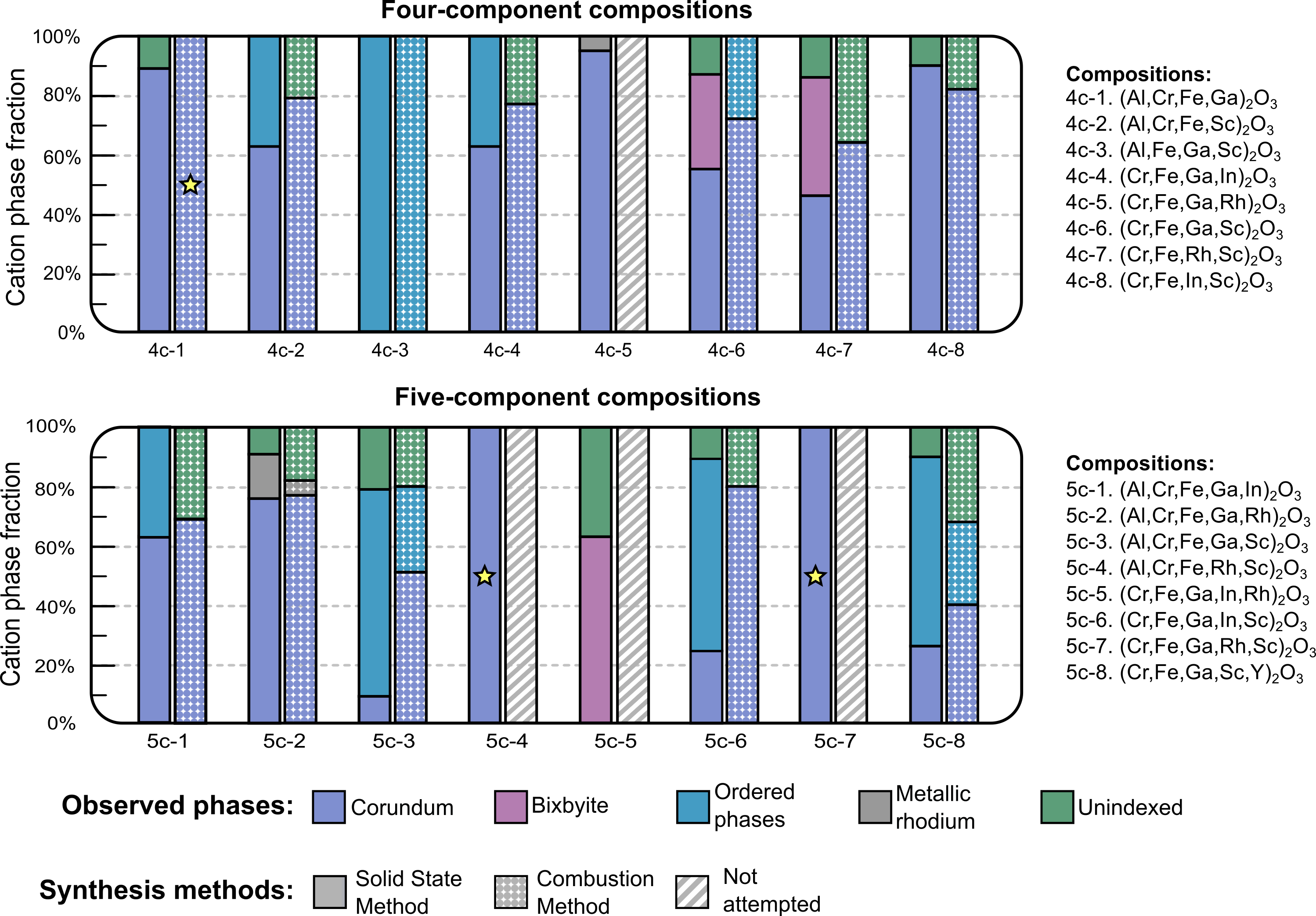}
 \caption{\textbf{Graphical representation of the synthesis outcomes for the 16 experimentally tested HEO compositions.} Cation phase fraction of different phases in all 16 compositions. The phases are represented with different colors, where blue represents the corundum structure(\textit{R}$\overline{3}$\textit{c}), purple represents the bixbyite structure (\textit{Ia}$\overline{3}$), cyan represents different ordered phases, including \textit{Pba}2, \textit{Pna}$2_1$, and \textit{Ia}$\overline{3}$\textit{d}, gray represents metallic rhodium and green represents the impurities that were estimated with numerical integration of the pXRD pattern. The solid bar represents the result of the solid state method and the bar with the dot pattern represents the result of the combustion method.}
\label{fig:bargraph}
\end{figure*}

\subsection{Overview}

%Leveraging the results of MLIPS, we can effectively reduce the exploration ground for novel trivalent HEOs into an experimentally resolvable scale. 
% This will depend on the previous sections structure
% Using the MLIPs results, we generated the plots shown in Fig. \ref{fig:MLIPs} to visualize the two descriptive parameters ($\Delta H_{MACE}$ and $\sigma_{E_i}$) calculated for both the corundum and bixbyite structures across all four and five component compounds. The figure consists of four plots: for each crystal structure, results are shown separately for the 210 four-component compositions and the 252 five-component compositions. Our selected candidates are highlighted in the plots using triangle markers, while the remaining compounds are shown as circles.

%  The selection process of the compositions to attempt was done using the numerical values of these parameters. We ordered them using the parameter $\Delta H_{MACE}$, and established the cutoffs of $\Delta H_{MACE} \leq 0.2 $ $\frac{\text{eV}}{\text{formula unit}} $ and $\sigma_{E_i}\leq0.075$ eV.

Employing the combined synthesizability metric, we selected 16 putative trivalent HEOs for experimental validation. %\textcolor{black}{A reader might wonder about the remaining possibilities within the shaded region that were not attempted?} 
These compositions included 8 four-component compounds and 8 five-component compounds, with a roughly equal distribution between predicted corundum and bixbyite phases. The selected compositions are presented in Table \ref{tab:Summaryattempts}, where they are ordered alphabetically by their elemental symbols. Each composition is also identified by a shortened label (\emph{e.g.} 4c-1 is the first listed four-component phase). It is interesting to note that there is a very uneven distribution of cations from our original subset of 10 represented in these 16 compositions. For example, Fe appears in every single composition, and Cr appears in all but one, while neither Bi nor Sb appears in any of the 16, which is understandable based on their lone pairs. Moreover, the smallest radii cations are heavily favored over the largest radii ones (see Fig.~\ref{fig:periodictable}(d)).

\begin{figure*}[htbp]
\centering
\includegraphics[width=\textwidth]{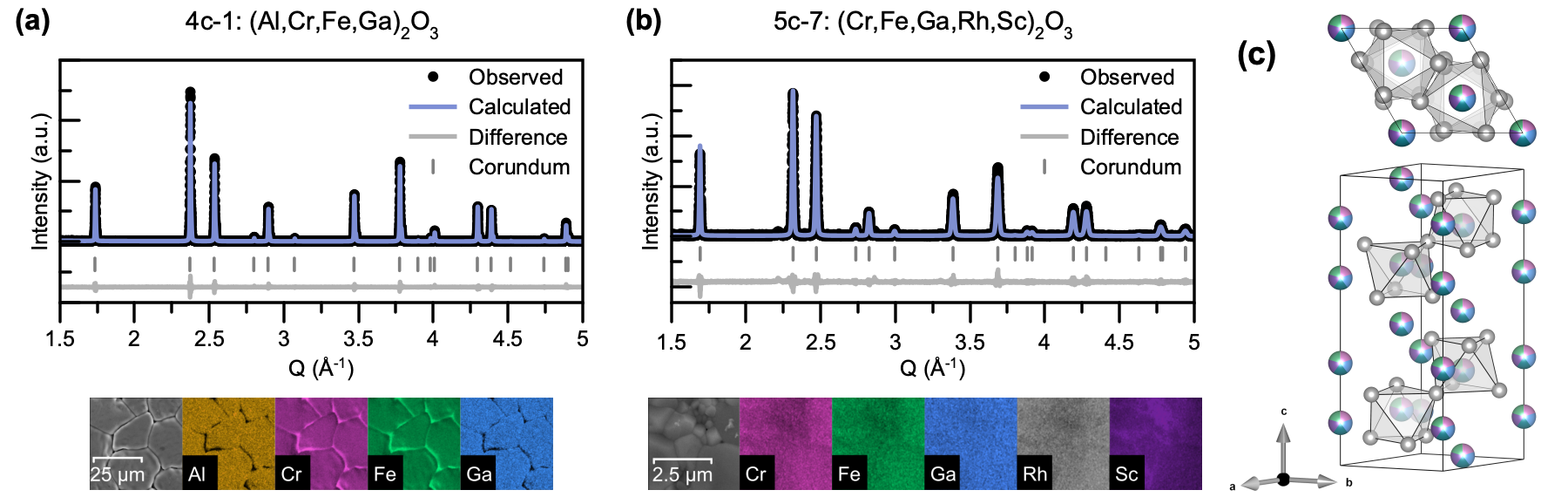}
 \caption{\textbf{Successful synthesis of novel corundum structured HEOs.} Rietveld refinement of the four-component composition \textbf{(a)} \ch{(Al,Cr,Fe,Ga)_2O_3} (4c-1) in the trigonal $R\overline{3}c$ corundum space group with goodness-of-fit parameters $R_{wp} =17.4\%$, $R_{exp} =15.2\%$ and $\chi^2 = 1.14$. The resulting phase formed with large domains and with no evident elemental segregation at the micron level as seen from the SEM image and EDS maps for the four cations. % Maybe consider adding label for EDS ad SEm images.
 Rietveld refinement for the five-component composition \textbf{(b)} \ch{(Cr,Fe,Ga,Rh,Sc)_2O_3} (5c-7) in the trigonal $R\overline{3}c$ corundum space group with goodness-of-fit parameters $R_{wp} = 24.0 \%$, $R_{exp} =18.0 \%$ and $\chi^2 = 1.34$. EDS maps show that the five cations are distributed homogenously at the micron scale. \textbf{(c)} Unit cell of a high entropy corundum structure in the hexagonal coordinate system with five different cations occupying the single cation crystallographic site.}
\label{fig:Successes}
\end{figure*}

Given that the outcome of HEO synthesis is known to be highly method- and condition-dependent~\cite{lin2020mechanochemical,aamlid2024effect,almishal_thermodynamics-inspired_2025}, we attempted the synthesis of most of these 16 compositions by both solid-state and combustion synthesis to leverage thermodynamic control and fast reaction kinetics, respectively~\cite{gonzalez-rivas_impact_of_synthesis}. Within each of these methods, we performed multiple iterations and optimizations with the goal of accessing the targeted high-entropy phase. The outcomes of these synthesis attempts are summarized in Table~\ref{tab:Summaryattempts} and presented graphically in Fig.~\ref{fig:bargraph}.   

The resulting methodology was able to identify three compositions (4c-1, 5c-4, and 5c-7) that formed into a phase pure HEO in the corundum structure, which are marked by stars in Fig.~\ref{fig:bargraph}. Notably, in all three cases, the phase pure HEO could be achieved exclusively by one of the two synthesis methods. 
%The first (4c-1) is notable for having the lowest enthalpy and entropy descriptors across all 462 compositions. Three of the four constituent cations occur in corundum structures at ambient conditions while Ga$_2$O$_3$ has a metastable corundum phase, ostensibly increasing the chances of forming a solid solution. The other two (5c-4 and 5c-7) both incorporate scandium, which is a cation that is not present in the corundum structure at ambient conditions. This could imply that entropy is playing a role in stabilizing the resulting structure. These five-component HEOs also both involve rhodium, which easily reduces at high temperatures, requiring tight control of synthesis conditions.  It is particularly interesting to note that neither of our phase pure five-component compositions include the same subset of constituents as the phase pure four-component phase, highlighting the delicate balance of factors influencing HEO formation.
The major pitfalls precluding the synthesis of an HEO were also identified from the synthesis results in Table~\ref{tab:Summaryattempts}. % using these 16 compositions as a representative sample. 
The most common pitfall was the presence of other phases competing for stability resulting in a mixture of phases, an outcome that was observed in 10 out of the 16 compositions. In 3 compositions the resulting phases strongly depended on the synthesis method, further calling attention to the importance of establishing the appropriate synthesis route. A second common pitfall was a change in oxidation state in one of the cations. This outcome was partially mitigated by our choice of cations, which were selected to have minimal competing oxidation states; nonetheless, rhodium is prone to reduction into its metal form at high temperature.  
%As a noble metal, rhodium tends to reduce to its metal form at high temperatures. This outcome can be mitigated using oxygen flow in solid state synthesis, but not completely for 3 of the 16 compositions. %As previously mentioned, this result can be mitigated using oxygen flow in solid state synthesis, but not completely for these 3 compositions. %Furthermore, the lack of thermodynamic control on combustion synthesis favored rhodium reduction. Therefore, most rhodium compositions were not pursued with it. 
Finally, in only 1 case (4c-3), we serendipitously discovered a novel phase pure material with significant cation ordering. In the following sections, we explore each of these outcomes in more detail. %This outcome was present only once with composition \ch{(Al,Fe,Ga,Sc)_2O_3}. The result was a novel single phase product with strong selectivity. This resulting phase was quite thermodynamically stable, as it resulted in both synthesis techniques, and it had been previously reported for the quaternary \ch{(Fe,Ga,Sc)_2O_3}.

\subsection{Successful HEO Synthesis}

The first phase pure corundum HEO, \ch{(Al,Cr,Fe,Ga)2O3} labeled as 4c-1, had the lowest enthalpy and entropy descriptors out of all 462 compositions considered and was predicted to form in the corundum structure. The favorable energetics can be understood as three of the four constituent cations occur in corundum structures at ambient conditions while Ga$_2$O$_3$ has a metastable corundum phase, ostensibly increasing the chances of forming a solid solution. Even still, this composition exhibited sensitivity to the synthesis method. The phase pure high entropy corundum could only be obtained via the combustion method, as shown by the Rietveld refinement in Fig.~\ref{fig:Successes}(a). The resulting phase, once sintered following combustion, had large (approximately 20 micron) domains with a homogeneous distribution of the cations, revealed by SEM and EDS.  

%It was confirmed via Rietveld refinement, with lattice parameters $a = 4.96097(2)$ \r{A} and $c = 13.47132(7)$ \r{A}. 

Attempts to synthesize \ch{(Al,Cr,Fe,Ga)2O3} (4c-1) via solid state reaction yielded a majority corundum product, even at the lowest sintering temperature (1100$^\circ$C), but with considerable impurity peaks in the pXRD, %on the pXRD pattern around 2.5 \r{A}$^{-1}$, 
as shown in Fig. S2 of the SM. Initially, these peaks decreased in intensity as temperature increased, but remained present and unchanged from 1300$^\circ$C to 1500$^\circ$C, accounting for approximately 10\% of the cation phase fraction (Fig.~\ref{fig:bargraph}). In order to explore if the limitation in phase purity arose from manual-grinding with a mortar and pestle, we performed the same growth with extensive ball-milling of the precursors prior to reaction. %The precursors were ground on a ball mill with a frequency of 35 HZ for 150 hours, and then sintered. 
The final product contained the same impurities, indicating that diffusion in the solid state cannot access this high entropy phase. It is interesting to note that a similar phase has been previously obtained with solid state synthesis through the addition of 2\% wt. \ch{TiO2} to act as a sintering agent, and enhancing the diffusion by generating charge center defects~\cite{WANG2024156998}. 

\begin{figure*}[htbp]
\centering
\includegraphics[width=\textwidth]{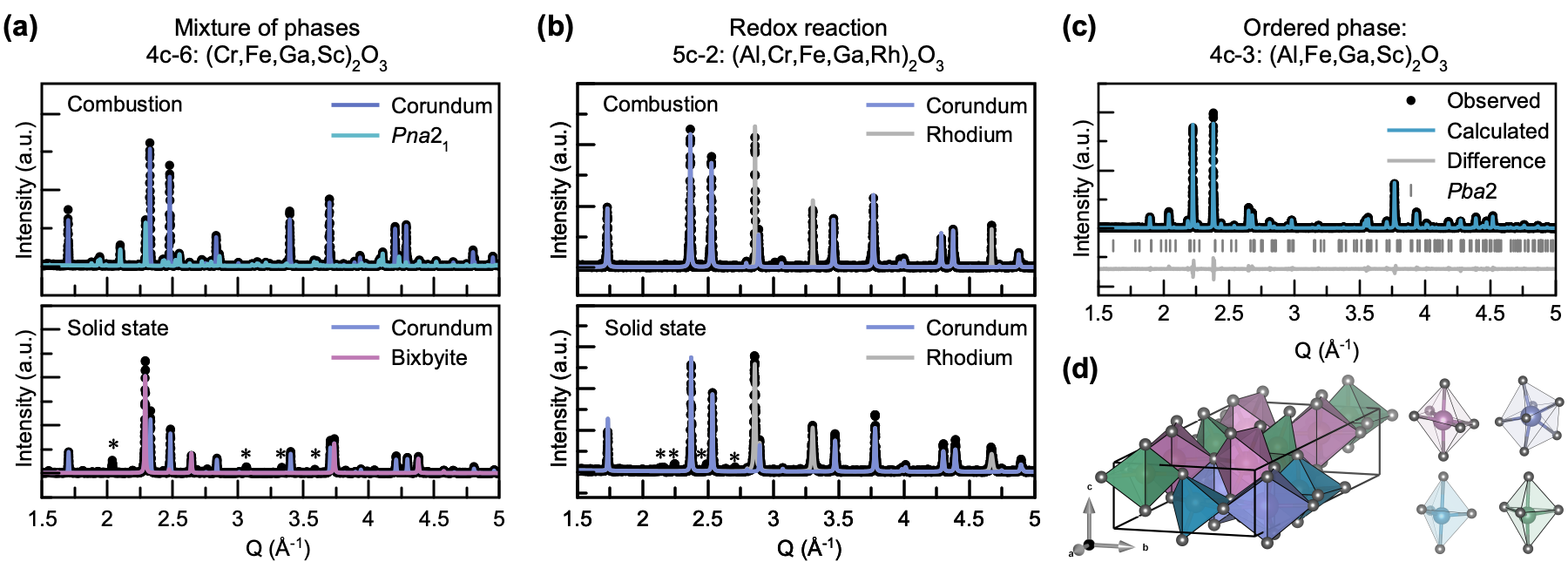}
 \caption{\textbf{Common pitfalls for the attempted synthesis of a novel HEO.} \textbf{(a)} The most common outcome across our 16 case studies was the formation of two or more competing phases, as exemplified by \ch{(Cr,Fe,Ga,Sc)2O3} (4c-6). This composition also reveals profound synthesis method-dependence. The phase-matched pXRD shows that combustion (top) yielded a mixture of corundum and an orthorhombic cation-ordered phase with space group $Pna2_1$ (no. 33), while solid state (bottom) favors a mixture of corundum and bixbyite phases. The asterisks mark an unindexed impurity phase. \textbf{(b)} Compositions containing rhodium were prone to reduction at elevated reaction temperatures, as observed for \ch{(Al,Cr,Fe,Ga,Rh)2O3} (5c-2) with both combustion (top) and solid state (bottom) synthesis in air. %In some cases, this reduction could be suppressed by flowing O$_2$ during solid-state synthesis. 
 \textbf{(c)} In only one case, \ch{(Al,Fe,Ga,Sc)2O3} (4c-3), we obtained a single-phase material with substantial cation ordering. This phase was stabilized by both solid-state and combustion synthesis. Rietveld refinement in the orthorhombic $Pba2$ space group gave goodness-of-fit parameters $R_{wp} = 17.62 \%$, $R_{exp} =14.85 \%$ and $\chi^2 = 1.19$. \textbf{(d)} Unit cell of the resulting crystal structure of \ch{(Al,Fe,Ga,Sc)2O3}, highlighting the four crystallographically distinct cation sites and their corresponding local environments.}
\label{fig:outcomes}
\end{figure*}

We also succeeded in synthesizing two five-component corundum HEOs, \ch{(Al,Cr,Fe,Rh,Sc)_2O3} and \ch{(Cr,Fe,Ga,Rh,Sc)_2O3}, which are labeled 5c-4 and 5c-7, respectively. A representative Rietveld refinement and elemental mapping for 5c-7 is shown in Fig.~\ref{fig:Successes}(b). The Rietveld refinement and refined crystal structures are found in Supporting information Figure S1 and Tables SI-SIII. 

These compositions are notable in particular for their inclusion of scandium, which is extremely insoluble in the corundum phase~\cite{prewitt1969c} and cannot be transformed to corundum even under extreme pressure~\cite{liu2009high}. %These components are not directly predictable with chemical intuition alone, mainly due to the presence of scandium, an element which not only is not present as corundum at ambient conditions but also it has not been reported to adopt it. % I looked on ICSD, only as a polar corundnum in ternearies like FeScO3 
In addition to scandium, composition 5c-7 also contains gallium, which in its binary oxide only transforms to a metastable corundum at extreme pressures~\cite{yan2008high}. Thus, 40\% of the cations in this composition are not enthalpically stable, making this phase a candidate for entropy stabilization~\cite{fracchia2024phase}. This composition is analogous to the prototype rock salt HEO (Mg,Ni,Co,Cu,Zn)O, in which both Cu and Zn are solubilized by the majority rock salt precursors~\cite{Rost2015,fracchia2022configurational}. %, with the latter having a high pressure rock salt polymorph. 
These five-component HEOs also exhibit thermodynamic restrictions in their synthesis due to rhodium, which will be discussed in Sect. \ref{sssec:oxidation} below.  %The very high instantaneous temperatures and uncontrolled atmospheric conditions of combustion synthesis are highly unfavorable for Rh containing compositions.  The reduction of rhodium required a control of the oxidizing atmosphere under which the reaction is performed. However, combustion synthesis is not an ideal method when requiring thermodynamic control. Attempts  with rhodium containing compositions quickly showed that trivalent rhodium could not be stabilized with combustion synthesis. Nonetheless, these compositions showed a strong preference for the corundum structure and the rhodium reduction was notably less prominent than in other compositions. In order, to assist the reaction we sintered them at higher temperatures (1200$^\circ$C) under flowing high purity \ch{O2}. Both of the compositions produced an almost phase pure results. Although tiny impurity peaks are still present at around 2.2 \r{A}$^{-1}$, these continuously decreased with temperature. %Both \ch{(Cr,Fe,Ga,Rh,Sc)_2O3} and \ch{(Al,Cr,Fe,Rh,Sc)_2O3} were confirmed to have a corundum structure using Rietveld refinement, with lattice parameters $a = 5.0848(1)$ \r{A} and $c = 13.7818(3)$ \r{A} and $a = 5.0578(2)$ \r{A} and $c = 13.7667(5)$ \r{A}, respectively.

\subsection{Other Outcomes}

Beyond the three phase-pure corundum HEOs, our 13 other experimentally validated compositions allowed us to form a comprehensive picture of HEO synthesis outcomes. These outcomes are categorized as mixtures of phases, oxidation state instabilities, and the formation of cation-ordered phases. Below, we discuss each of these outcomes using specific case studies to highlight key factors influencing the thermodynamic or kinetic barriers to forming an HEO.

\subsubsection{Mixture of phases}

In our experimental survey, the most common synthesis outcome was the formation of two or more competing stable phases, which occurred for 9 of the 16 compositions. In almost every case, the majority phase was corundum, which can be appreciated from inspection of Fig.~\ref{fig:bargraph} where corundum is represented by purple. The competing phases varied depending on the composition, and in 3 of the 16 (4c-6, 4c-7, and 5c-6), they even varied within the same composition with different synthesis methods. Although it is the ground state structure for three of our binary oxides, bixbyite was fairly underrepresented; it was the majority phase in just one case (5c-5) and appeared as a secondary phase in only 2 other compositions, and only when synthesized via solid-state. 

For many compositions, the competing phase was a complex oxide structure that we expect to have a significant degree of cation ordering. These phases are shown in blue in Fig.~\ref{fig:bargraph}. These structures include the garnet structure of \ch{Y5Al3O12} with space group \textit{Ia}$\overline{3}$\textit{d} (No. 230), the orthorhombic structure of the multiferroic \ch{AlFeO3} with space group \textit{Pna}$2_1$ (No. 33), the spinel structure of magnetite (\ch{Fe3O4}) %\textcolor{black}{Which composition had this? Fe3O4 has some 2+ cations which implies redox has occured!} \textcolor{blue}{Not necessarily, both gamma-Ga2O3 and gamma-Al2O3 are spinel structures with significant cation vacancies on the tetrahedral sites.} 
with space group \textit{Fd}$\overline{3}$\textit{m} (No. 227) and others. These structures each have multiple cation sublattices with distinct oxygen coordination numbers. The formation of these phases is necessarily driven by the preferential occupation of one or more sites by a set of specific cations, yielding a partially ordered structure. 

A useful case study is composition \ch{(Cr,Fe,Ga,Sc)2O3} (4c-6). The phase-matched pXRD patterns for the solid state and combustion synthesis products are presented in Fig.~\ref{fig:outcomes}(a), with the two methods yielding strikingly different outcomes. While corundum formed the majority phase in both cases, the secondary phase for combustion was the $Pna2_1$ structure of \ch{AlFeO3} while in solid state it was bixbyite. For the latter case, we can attribute the bixbyite phase, which accounts for approximately 30\% of the cation phase fraction, to the insolubility of Sc$_2$O$_3$ in the corundum phase. The $Pna2_1$ structure in the combustion product, meanwhile is more difficult to account for, as the AlFeO$_3$ structure, which is commonly described as an ordered derivative of the corundum structure, is not an obvious host for scandium. However, explorations of ScFeO$_3$ have shown that it may adopt this structure under high temperature and high-pressure (15 GPa) conditions~\cite{kawamoto2014room}. 

An additional unidentified impurity phase was also present in the solid-state sample, and these unindexed peaks are highlighted using asterisks. Most solid-state syntheses displayed unindexed peaks. These could not be identified, due to multiple factors: their intensities are much lower compared to the primary phase, other reflections could be overlapping with indexed peaks, and the complexity of the phase with 4 or 5 cations could lead to not yet reported ternary or quaternary phases. To estimate the proportion of these unindexed peaks, we have integrated the area of the calculated pattern with all the identified phases and compared it to the observed pattern, both of them with the background subtracted. This ratio, which assumes an equimolar cation distribution across all phases, provides a semi-quantitative estimate of the cation phase proportion. These unindexed phases are labeled in green in Fig.~\ref{fig:bargraph}.

% Not sure to add it but it's there
%Another remarkable example are compositions 4c-2, 4c-4, and 5c-1, these three compositions posses as usual corundum as the primary phase. However, they also posses the same secondary unidentified phase with peaks in positions  1.45 \r{A}$^{-1}$, 2.2 \r{A}$^{-1}$ and 2.3 \r{A}$^{-1}$ and similar intensities. Therefore, the phase exploration of novel HEOs could lead to discoveries of novel ternary or quaternary phases by recognizing patterns.

% \begin{itemize}
%     \item The most common outcome 9 out of 16 compositions.
%     \item Highlight corundum as a dominating phase across most compositions but with different competing space groups found in different compositions. Bixbyite was not very present. most common space groups were garnet, AlFeO3, some even depended on the synthesis method
%     \item Most solid state syntheses had unidentified phases. Therefore, we approximated the cation phase fraction with the proportion of the area of the unindexed peaks over the area of all the peaks.
%     \item Go over example with composition \ch{Cr,Fe,Ga,Sc)2O3}, the solid state synthesis ad the combustion both exhibit corundum, but with different competing phases, bixbyte in solid state and AlFeO3 in combustion. The solid state pXRD also highlights the unidentified phase peaks with asterisk. The relative intensity between them, the possible peak overlap, and the possible outcomes makes their phase identification difficult.
% \end{itemize}

\subsubsection{Oxidation state instability}\label{sssec:oxidation}

A second common outcome across our synthesis attempts was an undesired change in the oxidation state of one of our cations. As previously mentioned, this outcome was mainly observed with rhodium-containing compositions. As a noble metal, rhodium easily reduces to its metal state at high temperatures (>1100$^\circ$C). However, none of the 16 compositions considered here reacted into a single-phase product at temperatures below  1100$^\circ$C. Therefore, all rhodium compositions exhibited a redox reaction. A clear example is \ch{(Al,Cr,Fe,Ga,Rh)2O3} (5c-2), which is the compound with the lowest enthalpy and entropy descriptors of all five-component compositions. The phase-matched pXRD patterns shown in Fig.~\ref{fig:outcomes}(b) for combustion and solid state in air both display the peaks associated with rhodium metal, highlighted in gray. %at 2.85 \r{A}$^{-1}$ and 3.3 \r{A}$^{-1}$. 

Through control of reaction conditions, we partially mitigated this undesired reduction of rhodium. In the solid state synthesis, the rhodium-containing compositions were annealed under flowing high-purity \ch{O2} to favor rhodium oxidation. This approach was successful in the case of the previously mentioned successful corundum HEO syntheses of 5c-4 and 5c-7. However, in the case of 5c-2, the reaction under flowing O$_2$ had the effect of reducing the phase fraction of rhodium metal, but did not affect the proportion of the unindexed peaks highlighted with asterisks, as shown in Fig.~S2. %Notably, higher temperatures annealings under flowing high purity \ch{O2} appeared to favor a single phase in the two successful components where rhodium peaks had the lowest relative intensity out of all rhodium containing compositions. 
In combustion synthesis, we attempted to optimize the fuel by changing from glycine to ethylenediaminetetraacetic acid (EDTA), with the goal of producing a less volatile reaction. Even with these modifications and a low temperature post-combustion annealing (900$^\circ$C), the resulting phase exhibited both an incomplete reaction and substantial rhodium reduction. Therefore, we discarded combustion synthesis as a viable method for rhodium-containing compositions and it was not attempted in the case of 4c-5, 5c-4, 5c-5, and 5c-7. %The change of the cation's oxidation state is not within MLIPs capabilities, as a result candidates like 5c-2 can have lower synthesizability parameters.

It is worth emphasizing that a stable trivalent oxidation state was one of the key features we selected for in choosing our ensemble of possible cations. Therefore, we expect that under less tailored conditions, a redox outcome would be an even more common outcome from the attempted synthesis of an HEO. For example, manganese, which does exhibit a stable trivalent state, is prone to both reduction and oxidation when reacted with other transition metals, as can be observed in spinel HEOs. Furthermore, at present, redox compatibilities of reagents are not captured in current MLIP approaches. At a synthetic level, varying the reaction atmosphere is therefore a promising approach to stabilizing new HEOs, provided a compatible range can be identified~\cite{almishal_thermodynamics-inspired_2025}. There are, however, certain pairs of cations in specific oxidation states that cannot be simultaneously stabilized by conventional approaches~\cite{aamlid2024effect}.

\subsubsection{Novel materials discovery}

Finally, the least common outcome was the serendipitous discovery of a novel complex oxide, \ch{(Al,Fe,Ga,Sc)2O3} (4c-3), with a high degree of cation ordering. This phase forms in a structure belonging to space group \textit{Pba}2 (No. 32) and has been previously reported for the quaternary oxides \ch{In2Ga2Fe2O9} and \ch{Sc2Ga2Fe2O9}~\cite{AlFeGaSc}. In the case of \ch{(Al,Fe,Ga,Sc)2O3}, this phase was found to form by both synthesis methods and the Rietveld refinement for the combustion sample is shown in Fig.~\ref{fig:outcomes}(c). It also exhibits robust stability, occurring as a single phase via solid state at comparatively low temperatures (1200$^\circ$C). 

This orthorhombic structure, shown in Fig.~\ref{fig:outcomes}(d), contains four unique cation sites, two of which have 5-fold oxygen coordination while the other two sites are 6- and 7-fold coordinate, respectively. %The original compound sets scandium to be fully occupy the 7-coordinated site, while the remaining proportion of scandium is a solid solution with the other two cations in the other three crystallographic. 
Following the reported structure and considering scandium has the biggest ionic radius, we fixed scandium in the 7-coordinated environment and distributed the three remaining cations equimolarly between the three lower coordinate sites. The exact occupation and their of this sites cannot be conclusively obtained from pXRD. However, their site preferences were explored using a bond valence sum analysis, which is  shown in Table~SV in the supporting information. Our refinement showed excellent agreement with the measured diffraction pattern, and the tabulated refinement parameters are \textcolor{black}{provided in Table~SIV in the supporting information.} %with $GOF = 1.22$ and lattice parameters  $a = 18.6855(6)$\r{A} $a = 7.0480(2)$\r{A} and $c = 3.17405(8)$\r{A}. 

%However, we can not conclusively determine the occupation of each Wyckoff position. Notably, a solid solution of Al and Ga would be able to replicate the x-ray scattering strength of Fe and vice versa, due to their similar atomic numbers. \textcolor{black}{The resulting occupations showed some preference for some cations in specific sites; nonetheless, we performed the same refinement assuming equimolarity on all sites and the $\chi^2$ remained $1.22$. On the other hand, if we fix a single element, we do observe increases of the $GOF $ to  $\sim 1.7$.} \textcolor{blue}{I do not understand the two previous sentences. Do you mean we assume equimolarity of Al, Fe, and Ga on the three 5- and 6-fold coordinated sites? We also do not know what the GOF is for the refined case we report. Could this paragraph be moved to SI?} Therefore, the solid solution in the three sites is present, but in order to accurately measure the site occupancy, further studies similar to those presented in Ref. \cite{heo-spinel1} would be necessary. \textcolor{blue}{Since the two 5-coordinated sites are similarly coordinated but symmetrically inequivalent, resonant xrd (https://pubs.acs.org/doi/10.1021/acsomega.3c06500) or neutron powder diffraction should be preferred over xas.}

\section{Summary and outlook}

In this work, we have investigated the synthesizability of high entropy oxides of the form $A_2$O$_3$ from a selection of trivalent cations, demonstrating a complete materials discovery pipeline from machine-learning-accelerated high-throughput screening to experimental synthesis and structural characterization. Three new HEOs were discovered, namely the four-component \ch{(Al,Cr,Fe,Ga)2O3} and the two five-component \ch{(Cr,Fe,Ga,Rh,Sc)_2O3} and \ch{(Al,Cr,Fe,Rh,Sc)_2O3}. Additionally, \ch{(Al,Fe,Ga,Sc)2O3} was found to crystallize in a cation-ordered single phase. The five-component compounds are particularly interesting as they cannot be predicted based on chemical intuition from cation radii and phase of the elemental oxides alone, as well as requiring tight control of oxygen chemical potential during synthesis. We highlight \ch{(Cr,Fe,Ga,Rh,Sc)_2O3} as a notable candidate for entropy stabilization as neither Ga$_2$O$_3$ nor Sc$_2$O$_3$ favors a corundum structure in ambient conditions.

The barriers to synthesis for the remaining compounds primarily involve the formation of a mixture of phases and redox stability, and are not distinguishable by \textit{a priori} descriptor values alone. Control of the synthesis route is key to stabilizing the high entropy phase, evidenced by the differing results from solid state synthesis and combustion synthesis for the same compounds, which underscores the need for experimental verification of computationally predicted structures. Indeed, this has wider significance; recent large-scale computational materials discovery efforts have created databases of hundreds of thousands of predicted, new stable phases~\cite{merchant_scaling_2023}, yet evidence of successful synthesis of predicted phases has been limited to a handful of materials~\cite{szymanski_autonomous_2023}, and critical analysis of these syntheses has shown the difficulty in predicting and discovering novel materials that can be realized in the laboratory~\cite{leeman_challenges_2024,cheetham_artificial_2024}. 

In this context, the successful synthesis of 3 phase-pure HEOs from 16 candidates, screened from a compositional space of hundreds where no prior examples existed, shows the value of targeted screening through physics-informed computational descriptors that can handle the disorder inherent to HEOs. HEOs are far rarer than initially believed and this approach is accelerating their discovery by finding the ``needle in the haystack''. We expect that the integration of temperature- and pressure-dependent thermodynamic modeling within the MLIP-based screening framework will lead to even greater improvements in the predictive power of computational HEO discovery, allowing discovery hand-in-hand with tailored synthesis routes.

\section{Methods}

\subsection{Computational method}

The MACE-MP-0b2 medium density foundation model \cite{batatia2025foundation} was used to relax and calculate the descriptors of high entropy oxide supercells using the atomic simulation environment (ASE) ExpCellFilter~\cite{ase_paper} with a force convergence criteria of 0.05 eV/\AA. The approximately 1000 atom supercells are constructed from an initial unit cell. The cation sites are then randomly populated according to the HEO composition at the correct stoichiometry ratios (here, equimolarity).

From these relaxed structures, the descriptors are then calculated. The enthalpy descriptor is defined as
\begin{equation}
    \Delta H_{\text{MACE}} = E_{\text{HEO}}-\sum_{A} x_A E(A_2\text{O}_3)
    \label{eqn:deltaH1}
\end{equation}
where
\begin{equation}
    E(A_2\text{O}_3) = \frac{2}{x}E(A_x\text{O}_y)+(\frac{3}{2}-\frac{y}{x})E(\text{O}_2)
    \label{eqn:deltaH2}
\end{equation}
is the lowest enthalpy binary oxide of element $A$, accounting for the oxygen chemical potential at 0~K as calculated by MACE. All binary oxide crystal structures of the relevant cations were extracted from the Materials Project database~\cite{Jain2013,Horton2025}. These structures were relaxed with the lowest $E(A_2\text{O}_3)$ defining the enthalpy hull.

MLIPs such as MACE decompose the total energy of a system into a sum of individual atom energies, $E_{\text{total}} = \sum_{i} E_{i}$
where each $E_{i}$ is an individual atoms energy encoded from the full local chemical environment through many-body interactions. We exploit this property to define an entropy descriptor
\begin{equation}
   \sigma_{E_i} = \frac{\sum_{A}\sigma^{A}_{E_i}}{N_{\text{cations}}}
    \label{eqn:deltaH}
\end{equation}
where $N_{\text{cation}}$ is the number of cation species (4 or 5 here) and $\sigma^{A}_{E_{i}}$ is the standard deviation of each cation species' ($A$) individual atom energies, $E_{i}$, distribution
\begin{equation}
    \sigma^{\text{A}}_{E_{i}} = \sqrt{\frac{\sum_{i}|E^{\text{A}}_{i}-\overline{E^{A}_{i}}|^2}{N_{\text{A}}}} .
 \label{eq:entropy_descriptor_A}
\end{equation}
These descriptors are then combined, similarly to the mixed enthalpy-entropy descriptor (MEED) proposed by Dey \emph{et al.}~\cite{dey_mixed_2024}, to form a thermodynamic stability predictor, $\rho$, where
\begin{equation}
    \rho = \frac{1}{\sqrt{2}}\sqrt{\left(\frac{ \sigma_{E_{i}}}{w_{\sigma}} \right)^2 +\left(\frac{ \Delta H_{\text{MACE}}}{w_{H}} \right)^2} .
     \label{eq:MEED}
\end{equation}
Here $w_{H}$ and $w_{\sigma}$ are scaling factors that set the relative weight of each descriptor in the distance metric. In the absence of any previously synthesized trivalent HEOs against which to calibrate, we use the mean values of the $\Delta H_{\text{MACE}}$ and $\sigma_{E_i}$ distributions across the four-component corundum subspace.

\subsection{Experimental methods}

Polycrystalline samples were prepared via solid-state synthesis for all 16 compositions and combustion synthesis for 12 compositions, due to the aforementioned limitations with rhodium-containing compositions. The solid-state synthesis methodology consisted of mixing equimolar amounts of the constituent cations in their trivalent binary oxide form, $A_2$O$_3$. Considering the lowest melting point of the constituent trivalent oxides, the candidates were split into two sintering protocols. The first protocol was designed for compositions containing rhodium, 6 out of 16, which are limited by the decomposition temperature of \ch{Rh2O3} (1100$^\circ$C). The four or five binary trivalent oxides were mixed, pressed into a pellet, and fired at an initial sintering temperature of 800$^\circ$C, and then air-quenched. This procedure was then repeated in increasing temperature increments of 100$^\circ$C up to 1100$^\circ$C or until a single-phase product was obtained. The other 10 compositions were limited by the \ch{Fe2O3} melting point (1500$^\circ$C). These compositions followed the same workflow, but the initial sintering temperature was 1200$^\circ$C, and the maximum was 1500$^\circ$C. In all 16 solid-state syntheses, the weight of the pellet was measured before and after sintering to track any losses due to evaporation. For only (Al,Cr,Fe,Ga)$_2$O$_3$ (4c-1), the effect of pre-homogenization in solid state synthesis was explored with ball milling. The oxide precursors were ground with a frequency of 35 Hz for 120 hours, and then sintered at 600, 800, and 1000$^\circ$C for 24 hr.

The methodology for combustion synthesis was adapted from the protocol described in \cite{MAO2020165884}. Equimolar amounts of the four or five monometallic nitrates (\ch{$A$(NO_3)_3}) were dissolved in deionized water. The initial solution was then stirred for 30 minutes. The combustion reaction followed the following general form, where the first term represents the metallic salt solution:
\begin{equation}
    \begin{split}
    2(A,B,C,D) (\text{NO}_3)_3 + \frac{10}{3} \phi \text{C}_2\text{H}_5\text{NO}_2  + \frac{15}{2} (\phi -1)\text{O}_2 \\ \xrightarrow{} 
    (A,B,C,D)_2\text{O}_3  + \frac{25}{3} \phi \text{H}_2\text{O} + \frac{5\phi +9}{3}\text{N}_2 + \frac{20}{3} \phi \text{CO}_2
    \end{split}
    \label{eqn:balancing equation comb}
\end{equation}

The appropriate amount of fuel, glycine (\ch{C_2H_5NO_2}), was added and stirred for another 30 minutes. The oxidants-to-reducers ratio $\phi$ was fixed to 1. The solution was then heated using a hot plate set to 160$^\circ$C to slowly remove the solvent and form a xerogel. The hot plate was then set to 350$^\circ$C to trigger combustion of the xerogel. The product was then placed in a preheated furnace at 1300$^\circ$C for 12 hours and subsequently quenched. This step removed any remaining organic products and improved the sample's overall crystallinity. 

For rhodium-containing compositions, the method was initially optimized for the growth of composition 5c-2. The rhodium cation can only be incorporated in the oxide form, due to the lack of a commercially available rhodium salt. As rhodium is prone to reduction, and present as a powder immobilized in the xerogel, the reaction was designed to be less violent than in the glycine case. To this end, two longer-chain organic molecules were tested -- EDTA (\ch{C10H16N2O8}) and citric acid (\ch{C_6H_8O_7}). Samples synthesized using these fuels where then post-growth annealed at 900$^\circ$C, to avoid reduction of the rhodium cations. Both resulting samples still contained significant amounts of reduced rhodium metal, although these impurities were less prominent in the EDTA product. %Therefore, combustion could not synthesized rhodium compositions without reduction and no further combustion synthesis was pursued in rhodium compounds.

The structural characterization of the synthesized compositions was resolved using powder x-ray diffraction (pXRD). Measurements were performed on a Bruker D8 Advance diffractometer using a Cu x-ray source with monochromated $\lambda_{K\alpha_1}$ = 1.5418 \r{A} in Bragg-Brentano geometry at room temperature. For solid-state syntheses, a pXRD pattern was measured after each sintering to observe the structural evolution as a function of sintering temperature, and for combustion syntheses, the patterns were only collected for the final product. Once the phase or phases were identified for each composition, Rietveld refinements were performed using TOPAS~\cite{coelho2018topas}. Refined quantities include the background, sample displacement, lattice constants, atomic positions, site occupancies, modified peak shape, strain broadening, and thermal parameters. 

The resulting phase fractions in Fig.~\ref{fig:bargraph} were obtained from the weight fractions generated by TOPAS, and converted into the molar fractions of the identified phases. For each composition with unindexed peaks, the unknown cation phase fraction was estimated by computing the area of the indexed peaks obtained with the refinement and the total area of the measured pXRD with the background subtracted. The final proportion of unknown impurities is given then by $1-\frac{\text{Indexed}}{\text{Total}}$. This calculation gives an approximate estimate of these secondary (or tertiary) phases, as it assumes both the indexed and unindexed structures have the same scattering strength and crystallinity.

\section*{Data availability}

The data that support the findings of this study are available from the corresponding author upon reasonable request.

\section{Acknowledgments}

The authors thank Ola Gjønnes Grendal for indexing the lattice parameters of \ch{(Al,Fe,Ga,Sc)2O3}. This work was supported by the NSERC-NSF Designing Materials to Revolutionize and Engineer our Future (DMREF) Alliance Grant in partnership with DMREF Grant Number DMR2523217. Additional support was provided by the Natural Sciences and Engineering Research Council of Canada (NSERC) and the Canadian Institute for Advanced Research (CIFAR). AMH was supported by the Killam Accelerator Research Fellowship. OAD was supported by the UKRI Guarantee Marie Skłodowska-Curie grant administered through the Engineering and Physical Sciences Research Council (EP/X034429/1).

\section*{Author contributions}

The computational screening was performed by O.A.D. with input from S.S.A. and J.R. Experimental validation was performed by A.A.M., with support from M.U.G.R. on the combustion methodology and K.T. on the ball-milled sample. Analysis of the experimental outcomes was performed by A.A.M., S.S.A., D.S., and A.M.H. The manuscript was written by A.A.M., O.A.D., S.S.A., and A.M.H. with input from all co-authors.

%\section*{Competing interests}
%The authors declare no competing interests.

%\bibliographystyle{natbib}
%\bibliographystyle{apsrev4-2}
%\bibliographystyle{apsrmp4-2.bst}
%\bibliographystyle{unsrtnat}

\bibliography{refs}% Produces the bibliography via BibTeX.

\end{document}